\documentclass[sigconf]{acmart}
\usepackage{pbalance}
\usepackage{booktabs}
\usepackage{tabularx}
\usepackage{rotating}
\usepackage{threeparttable}

\AtBeginDocument{%
  }

\title[]{CODENS: Transforming Code Changes into Living, Accessible, and Queryable Documentation}

\author{Abdelhak Kelious}
\affiliation{%
  \institution{CAPSENS}
  \city{Paris}
  \country{France}}
\email{Abdelhak.Kelious@capsens.eu}

\author{Chyrine Tahri}
\authornote{Work conducted while at Capsens.}
\affiliation{%
  \institution{CAPSENS}
  \city{Paris}
  \country{France}}
\email{chyrine@capsens.eu}

\author{Eliot Bardet}
\affiliation{%
  \institution{CAPSENS}
  \city{Paris}
  \country{France}}
\email{eliot@capsens.eu}

\renewcommand{\shortauthors}{Kelious et al.}

\setcopyright{cc}
\setcctype{by}

\acmConference[DocEng '26]{Proceedings of the 2026 ACM Symposium on Document Engineering}{August 25--28, 2026}{Fribourg, Switzerland}
\acmBooktitle{Proceedings of the 2026 ACM Symposium on Document Engineering (DocEng '26), August 25--28, 2026, Fribourg, Switzerland}

\acmYear{2026}
\copyrightyear{2026}

\acmDOI{10.1145/3820755.3832807}
\acmISBN{979-8-4007-2786-3/2026/08}

\begin{document}

\begin{abstract}
Maintaining up-to-date code documentation is difficult in fast-moving repositories because design knowledge is scattered across source files and pull requests. We present \textbf{CODENS}, a system that turns pull requests into living, accessible, and queryable documentation for production codebases. CODENS incrementally builds a typed software knowledge graph from pull requests, enriches components through schema-driven semantic extraction, derives typed relations between them, and exposes the resulting knowledge through three retrieval modes, including agent-guided graph traversal for repository-level question answering. The system also preserves semantic change history across pull requests and integrates both answer-quality and operational evaluation metrics. We evaluate CODENS on a client Ruby on Rails project in production. Results show that CODENS produces highly relevant and well-grounded answers, while qualitative feedback highlights a remaining challenge in concise, documentation-oriented synthesis. 
\end{abstract}

\ccsdesc[500]{Software and its engineering~Software maintenance tools}
\ccsdesc[500]{Information systems~Information retrieval}

\keywords{Knowledge graph, Code documentation, Retrieval-augmented generation, Large language models, Pull request analysis, Graph-based RAG, Software maintenance, Agentic retrieval, Neo4j, Ruby on Rails}

\maketitle

\begin{figure*}[t]
\centering
\includegraphics[width=0.99\textwidth]{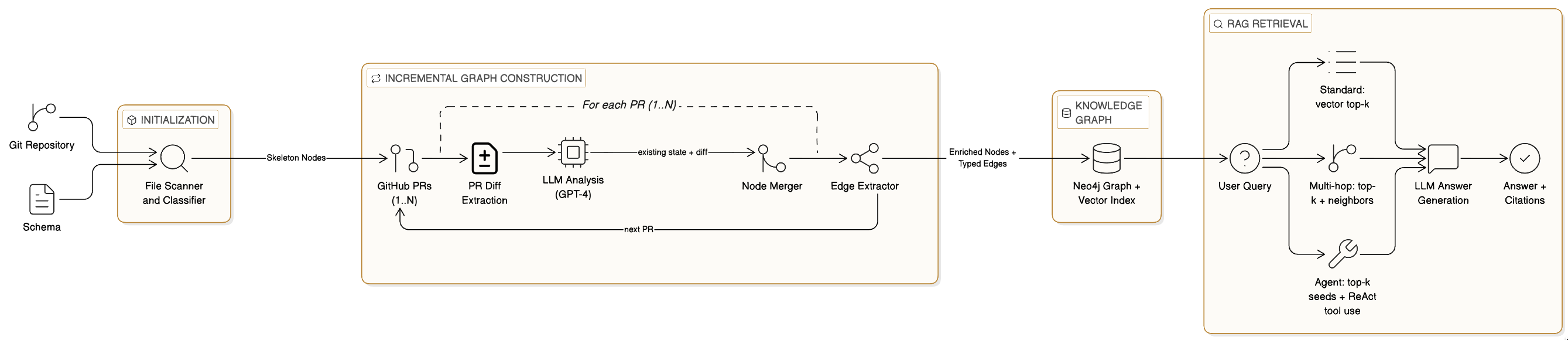}
\caption{\small CODENS pipeline overview. The system processes pull requests incrementally to build a semantic knowledge graph.}
\label{fig:architecture}
\end{figure*}

\section{Introduction}

Keeping software documentation up to date is difficult in fast-moving repositories. 
Design knowledge is often scattered across source files, pull requests, code reviews, and informal developer discussions. 
As a result, documentation quickly becomes incomplete or obsolete, especially in legacy systems where the rationale behind past changes is no longer easily accessible. 
This problem is particularly important for repository-scale code understanding, where answering a developer question often requires reasoning across multiple files, components, and historical changes.

Recent work has shown that structured representations of software artifacts can support repository-level understanding. 
Static-analysis-based software knowledge graphs such as \emph{KG4Py} construct graphs from source code using concrete syntax tree analysis to support semantic code search~\cite{Liang2022KG4Py}, while more recent systems such as \emph{CodexGraph} connect LLM agents to graph databases extracted from repositories for repository-level software engineering tasks~\cite{Liu2025CodexGraph}. 
In parallel, retrieval-augmented generation for code has moved beyond flat chunk retrieval: \emph{RepoCoder} introduced iterative retrieval-generation over repository artifacts~\cite{Zhang2023RepoCoder}, and \emph{GraphCoder} showed that a code context graph capturing structural dependencies improves repository-level context retrieval~\cite{Liu2024GraphCoder}. 
However, these approaches typically build or query knowledge from a repository snapshot, and do not explicitly turn the pull-request history of a legacy project into cumulative documentation knowledge.

We present \textbf{CODENS}, a tool that transforms code changes into living, accessible, and queryable documentation. 
CODENS first constructs an initial software knowledge graph for an existing Ruby on Rails codebase by scanning the repository and creating typed skeleton nodes from the framework's architectural conventions, such as controllers, models, views, services, policies, jobs, and tests. 
It then replays pull requests chronologically: for each changed file, CODENS injects both the current semantic state of the corresponding node and the pull-request diff into an LLM, extracts schema-driven attributes and relations, and incrementally merges them into the graph. 
In this way, CODENS turns both legacy change history and future pull requests into a persistent memory layer for documentation and question answering. 
In this work, CODENS is instantiated for Ruby on Rails, leveraging the framework's architectural conventions through a schema that could be adapted to other frameworks with similarly explicit structures.

CODENS differs from prior work on code summarization and change explanation in its objective. 
Prior approaches mainly generate summaries at the level of functions, files, or individual changes. 
For example, \emph{CodeT5} provides a unified framework for code understanding and generation~\cite{Wang2021CodeT5}, while Sun et al.\ show that LLMs improve source code summarization but still mainly operate at local granularity~\cite{Sun2025CodeSummarization}. 
At the change level, Li et al.\ show that \emph{diff alone is not enough} for high-quality commit message generation and that broader repository context is necessary~\cite{Li2024OnlyDiff}. 
CODENS builds on this intuition, but instead of producing one-shot textual explanations, it continuously updates a structured project memory with semantic attributes such as purpose, behavioral flow, business logic, and typed inter-component relations.

CODENS also combines graph-based and agentic retrieval for documentation-oriented question answering. 
Microsoft \emph{GraphRAG} showed that graph indexes can improve question answering over large corpora by organizing evidence through graph structure and community-level summaries~\cite{Edge2024GraphRAG}. 
More generally, \emph{ReAct} established the benefit of interleaving reasoning and tool use~\cite{Yao2023ReAct}, and \emph{SWE-agent} showed that software engineering agents strongly benefit from carefully designed action interfaces~\cite{Yang2024SWEAgent}. 
CODENS applies these ideas to a typed software graph and exposes three retrieval modes: standard vector retrieval, graph-expanded multi-hop retrieval, and agent-guided traversal through graph-specific tools such as \texttt{GET\_NODE}, \texttt{GET\_NEIGHBORS}, \texttt{GET\_RELATIONS}, \texttt{CYPHER}, and \texttt{ANSWER}. The main contribution of this work is a framework-aware pipeline for constructing and maintaining a typed software knowledge graph from a production Ruby on Rails codebase. The proposed system incrementally processes pull requests, enriches graph nodes with LLM-extracted semantic attributes, and preserves change history across software evolution. On top of this graph, CODENS provides a documentation-oriented chatbot that combines vector retrieval, graph-expanded multi-hop retrieval, and agentic traversal through graph-specific tools. We evaluate the approach on a production project using human assessment, RAG metrics, and operational indicators such as latency, token usage, cost, and estimated carbon footprint.

\section{Tool Architecture}

We build, validate, and evaluate CODENS on a production Ruby on Rails application developed by one of our clients---a web platform comprising over 1,700~source files organized following the standard Rails MVC \footnote{MVC stands for Model--View--Controller, a software architectural pattern that separates data models, user interface views, and request-handling controllers.} architecture (models, views, controllers, services, policies, jobs, etc.). The repository has an active development history with hundreds of pull requests spanning several months. \textit{For confidentiality reasons, we do not disclose the name or business domain of the application }. Figure~\ref{fig:architecture} presents the complete architecture of CODENS, organized into four functional blocks: initialization, incremental graph construction, knowledge base storage, and multi-mode RAG querying.

\subsection{Initialization}

The system takes as input a Git repository and a JSON schema defining the attributes to extract per component type. In the Rails framework, each source file has a well-defined role determined by its directory: files in \texttt{app/controllers/} handle HTTP requests, files in \texttt{app/models/} define data and business logic, files in \texttt{app/views/} render HTML, and so on. CODENS leverages these conventions: a scanner classifies each source file into one of 17~component types (Controller, Model, View, Service, Policy, Spec, Locale, Validator, Job, Worker, etc.) and creates a \textbf{skeleton node} containing only structural metadata---a unique \texttt{node\_id} derived from the file path (e.g., \texttt{app.controllers.users\_controller}), the component type, and the source path. No semantic content is extracted at this stage. On our evaluation project, this step produces \textbf{1,739~skeleton nodes}.

\subsection{Incremental Graph Construction}

This block is the core of CODENS. It processes pull requests chronologically, enriching skeleton nodes with semantic attributes through five steps per~PR:

\textbf{(1)~Diff extraction:} the system fetches modified files and their patches from the GitHub API, retaining only files matching existing nodes.

\textbf{(2)~State injection.} This step is a key design choice. 
In CODENS, the \emph{state} of a node denotes the current semantic representation already stored in the graph before processing the current pull request. 
It includes the node type and path, previously extracted semantic attributes such as \texttt{purpose}, \texttt{behavioral\_flow}, \texttt{business\_logic}, and \texttt{invoked\_models}, as well as known relationships and provenance metadata. 
If such a state already exists, CODENS injects it into the LLM prompt together with the new PR diff. 
Without this state, each PR would produce a description based solely on its local diff, causing earlier knowledge to be lost. With state injection, the LLM is instructed to update the existing representation rather than replace it. For example, if a controller node already describes two actions (\texttt{index}, \texttt{show}) and the current PR adds a third action (\texttt{create}), the prompt contains both the previous node state and the new diff. 
The LLM then produces an updated \texttt{behavioral\_flow} that integrates all three actions, instead of describing only the newly added one.

\textbf{(3)~LLM analysis:} each file is submitted to GPT-4 with the patch, the schema fields for its component type, the existing node state, and merge instructions. The LLM returns structured JSON attribute changes, validated against the schema.

\textbf{(4)~Node merging:} changes are applied using a differentiated strategy. Scalar narrative fields (e.g., \texttt{behavioral\_flow}, \texttt{purpose}) are replaced, with the previous value archived in a per-PR history. List fields (e.g., \texttt{invoked\_models}, \texttt{relationships}) are merged preserving uniqueness. Each node also tracks provenance metadata: the first PR that introduced it, the last PR that modified it, and the full ordered list of PRs that touched it.

\textbf{(5)~Edge extraction:} after all PRs, an edge extractor creates typed relationships between nodes using two strategies: (a)~schema-driven, reading relational fields from enriched nodes and resolving them to target nodes via a name index; (b)~cue-based, applying regex patterns from a taxonomy file on source code. Edges are deduplicated by \texttt{(type, source, target)}. On our evaluation project, this produces \textbf{622~unique edges} across 11~relationship types (e.g., \texttt{invokes\_model}, \texttt{renders\_view}, \texttt{tests\_component}).

\subsection{Knowledge Base}

Enriched nodes and typed edges are imported into \textbf{Neo4j}, a native graph database. For each node, a text representation is composed from its key semantic attributes---prioritizing \texttt{behavioral\_flow}, which captures the step-by-step behavior of the component---and embedded using sentence-transformers (\texttt{all-mpnet-base-v2}) \footnote{Sentence-Transformers is a Python library for generating dense vector representations of sentences and documents, commonly used for semantic similarity, clustering, and retrieval tasks.}. The resulting vectors are stored as node properties in Neo4j and indexed for cosine similarity search, yielding a knowledge graph that supports both relational traversal and semantic search.

\subsection{Multi-Mode RAG Querying}

The system exposes a Streamlit chatbot with three retrieval modes of increasing depth, The chatbot makes the knowledge graph directly accessible to developers: instead of manually inspecting graph nodes or writing Cypher queries, users can ask natural-language questions about features, flows, business rules, or component interactions and receive grounded answers synthesized from the graph.

\textbf{Standard:} the user question is embedded and the top-$k$ \footnote{Top-$k$ denotes the $k$ highest-scoring nodes according to cosine similarity between the embedded user question and the node embeddings.} most similar nodes are retrieved via vector search. Their text is concatenated and sent as context to the LLM, which generates a response. This mode is fast but limited to nodes whose text is directly similar to the question---it does not follow graph relationships.

\textbf{Multi-hop:} starting from the same top-$k$ seeds, the system traverses graph edges to retrieve neighboring nodes, filters them by cosine similarity with the question, and adds the most relevant ones to the context. This captures one or two hops of relationships but the expansion is automatic and undirected---every relationship type is followed equally.

\textbf{Agent:} the top-$k$ seeds are presented to a ReAct agent~\cite{Yao2023ReAct} that \emph{decides} which parts of the graph to explore. The agent is equipped with five tools (Table~\ref{tab:agent-tools}) and follows a reason-act-observe loop: at each step, it examines its current knowledge, selects a tool call, incorporates the result, and iterates until it can produce a final answer. Unlike multi-hop, the agent can selectively follow specific relationship types, skip irrelevant branches, and issue arbitrary Cypher queries when needed. This mode is the most expensive in tokens but produces the most complete answers, as graph exploration is guided by the LLM's reasoning rather than a fixed expansion strategy.

\begin{table}[h]
\centering
\small
\begin{tabular}{ll}
\toprule
\textbf{Tool} & \textbf{Description} \\
\midrule
\texttt{GET\_NODE(node\_id)} & Read all attributes of a node \\
\texttt{GET\_NEIGHBORS(node\_id, ...)} & Retrieve neighbors, filterable by type \\
\texttt{GET\_RELATIONS(node\_id)} & List relationship types with counts \\
\texttt{CYPHER(query)} & Execute an arbitrary Cypher query \\
\texttt{ANSWER(text)} & Provide the final answer \\
\bottomrule
\end{tabular}
\caption{Tools available to the CODENS agent during graph exploration.}
\label{tab:agent-tools}
\end{table}

\subsection{Execution Metrics}

For each query, the chatbot displays a metrics panel reporting execution duration, model used, retrieval mode, number of nodes retrieved, input and output tokens, number of LLM calls, estimated monetary cost based on the model's per-token pricing, and estimated carbon footprint based on average energy consumption per token and the electrical grid's carbon intensity.




\begin{table*}[!t]
\centering

\caption{\small Per-query evaluation results. Questions are anonymized as Q1--Q11.}
\label{tab:evaluation_metrics}
\tiny
\setlength{\tabcolsep}{2pt}
\renewcommand{\arraystretch}{0.90}
\begin{tabular*}{\textwidth}{@{\extracolsep{\fill}}lccccccccrrrrr@{}}
\toprule
\textbf{Q}
& \multicolumn{4}{c}{\textbf{Human}}
& \multicolumn{4}{c}{\textbf{Automatic}}
& \multicolumn{5}{c}{\textbf{Cost \& Performance}} \\
\cmidrule(lr){2-5} \cmidrule(lr){6-9} \cmidrule(lr){10-14}
& \textbf{Rel.} & \textbf{Comp.} & \textbf{Doc.} & \textbf{Avg}
& \textbf{Ctx.} & \textbf{Faith.} & \textbf{Ans.} & \textbf{Avg}
& \textbf{Tok.} & \textbf{Cost} & \textbf{Time} & \textbf{Steps} & \textbf{CO$_2$} \\
\midrule
Q1  & 4 & 5 & 5 & 4.67 & 1.00 & 1.00 & 0.944 & 0.981 & 31,640 & .067 & 12.6 & 8  & 6.17 \\
Q2  & 5 & 5 & 5 & 5.00 & 1.00 & 1.00 & 0.967 & 0.989 & 15,433 & .036 & 14.2 & 7  & 3.01 \\
Q3  & 5 & 5 & 5 & 5.00 & 1.00 & 1.00 & 0.877 & 0.959 & 14,812 & .037 & 20.6 & 4  & 2.89 \\
Q4  & 4 & 4 & 5 & 4.33 & 1.00 & 1.00 & 0.942 & 0.981 & 41,150 & .091 & 30.0 & 16 & 8.02 \\
Q5  & 4 & 5 & 5 & 4.67 & 1.00 & 1.00 & 0.906 & 0.969 & 42,363 & .091 & 21.8 & 12 & 8.26 \\
Q6  & 3 & 3 & 5 & 3.67 & 1.00 & 1.00 & 0.921 & 0.974 & 40,557 & .089 & 25.7 & 11 & 7.91 \\
Q7  & 4 & 5 & 5 & 4.67 & 1.00 & 1.00 & 0.948 & 0.983 & 18,664 & .044 & 23.7 & 7  & 3.64 \\
Q8  & 3 & 5 & 5 & 4.33 & 1.00 & 1.00 & 0.956 & 0.985 & 50,788 & .108 & 27.8 & 10 & 9.90 \\
Q9  & 3 & 5 & 5 & 4.33 & 1.00 & 1.00 & 0.945 & 0.982 & 13,609 & .030 & 11.8 & 6  & 2.65 \\
Q10 & 5 & 3 & 5 & 4.33 & 1.00 & 1.00 & 0.925 & 0.975 & 38,836 & .087 & 22.6 & 8  & 7.57 \\
Q11 & 5 & 4 & 4 & 4.33 & 1.00 & 1.00 & 0.985 & 0.995 & 64,419 & .140 & 38.5 & 24 & 12.56 \\
\bottomrule
\end{tabular*}

\vspace{-0.5em}
\begin{flushleft}
\tiny
\textit{Note.} Rel.=relevance; Comp.=completeness; Doc.=document relevance; Ctx.=context precision; Faith.=faithfulness; Ans.=answer relevancy. Human scores: 1--5; automatic metrics: 0--1; time in seconds; CO$_2$ in grams.
\end{flushleft}
\vspace{-1em}
\end{table*}

\begin{table*}[!t]
\centering
\caption{\small Qualitative feedback by query. Questions are anonymized as Q1--Q11.}
\label{tab:qualitative_feedback}
\tiny
\setlength{\tabcolsep}{12pt}
\renewcommand{\arraystretch}{0.80}
\begin{tabularx}{\textwidth}{@{}lX@{}}
\toprule
\textbf{Question} & \textbf{Feedback} \\
\midrule
Q1 & Very good, but too technical and overly focused on point-by-point code citation. \\
Q2 & Perfect; it would be even better to mention file names rather than only graph nodes. \\
Q3 & Good overall. The question was somewhat ambiguous, so obtaining a fully clear answer was difficult; the response is acceptable. \\
Q4 & Good, but again very long. It should be more synthetic, although it is quite complete. It could also infer that the user wants to know on which page this flow is located. \\
Q5 & The answer could be clearer and should interpret the code and steps more instead of simply listing them; nevertheless, it answers the question well. \\
Q6 & The answer is vague and long, and in this case incomplete given the question. A full end-to-end user flow was expected, including the page to visit and the button to click. \\
Q7 & Still lacks synthesis and remains hard to read, but the answer is very complete. \\
Q8 & The answer is very complete but difficult to read. It struggles to synthesize information clearly. It should cite the code less line-by-line and provide more explanatory synthesis, even if the final answer is shorter. \\
Q9 & It correctly identifies that there are two policies, but the final answer focuses on only one of them without knowing what the user actually wanted. The question was admittedly ambiguous. \\
Q10 & The question is very vague, so a fully complete answer is not expected. The response is fairly good because it scans broadly to answer. \\
Q11 & No feedback. \\
\bottomrule
\end{tabularx}

\vspace{-0.5em}
\begin{flushleft}
\tiny

\end{flushleft}
\vspace{-1em}
\end{table*}
\section{Evaluation Results}

Tables~\ref{tab:evaluation_metrics} and~\ref{tab:qualitative_feedback} report an initial user-oriented evaluation of CODENS.
We focus on the \emph{agent mode} because it is the most expressive retrieval mode: it can inspect nodes, follow specific graph relations, and issue Cypher queries when needed.
This makes it suitable for documentation questions that require reasoning across multiple components. For this paper, we restricted the study to 11 anonymized questions derived from a client project in production. Both the questions and the answer quality assessments were reviewed by the lead developer of that project. Quantitative results are strong overall: average human scores reach 4.09/5 for relevance, 4.45/5 for completeness, and 4.91/5 for document relevance. Automatic metrics are also consistently high, with perfect context precision and faithfulness for all queries and an average answer relevancy of 0.94. These results indicate that CODENS generally retrieves appropriate evidence and remains well grounded in that evidence.

The qualitative feedback provides a more nuanced view of these results. While the retrieved evidence was generally considered relevant, several answers were perceived as too detailed or too code-centric for documentation use. The evaluator expected more synthesis, clearer explanations of functional flows, and more user-oriented references such as files, pages, or end-to-end UI paths. This indicates that the main improvement opportunity lies less in retrieval quality than in answer presentation: CODENS should better adapt its responses to the level of abstraction expected from living documentation. Given that this initial study covers 11 questions on a single industrial codebase, we interpret the results as exploratory but encouraging.

\section{Conclusion}

CODENS explores how software knowledge graphs can support living documentation by creating, enriching, and preserving repository knowledge over time. 
Starting from an existing Ruby on Rails codebase, CODENS constructs a typed graph from framework-level conventions, then enriches it by replaying pull requests and extracting semantic information from code changes. 
The resulting graph provides a persistent documentation layer that evolves with the repository and can be queried through multiple retrieval modes.

Our initial evaluation on 11 questions from a single industrial codebase suggests that CODENS can retrieve relevant evidence and generate well-grounded answers, with high document relevance, context precision, and faithfulness scores. 
However, these results are exploratory and should not be interpreted as a comprehensive assessment. 
A broader study across more projects, questions, users, and retrieval modes is needed to assess generality.

The qualitative feedback indicates that future improvements should focus on answer formulation as much as retrieval quality. 
Several answers were considered too verbose or too code-centric for documentation use, suggesting the need for more concise, synthesized, and user-oriented responses. 
Finally, while preliminary observations suggest possible operational advantages in cost, token usage, and response time, a systematic comparative benchmark remains future work.

\bibliographystyle{ACM-Reference-Format}
\bibliography{sample-base}

\end{document}